\documentclass[
aps,%
12pt,%
final,%
notitlepage,%
oneside,%
onecolumn,%
nobibnotes,%
nofootinbib,
superscriptaddress,%
noshowpacs,%
centertags]%
{revtex4}
\usepackage{epsfig,color}
\usepackage{epsfig}
\usepackage{lscape}
\usepackage{amsmath}
\usepackage{amssymb}
\usepackage{graphicx}
\usepackage{graphics}
\usepackage{subfigure}

\def\bea{\begin{eqnarray}}
\def\eea{\end{eqnarray}}
\def\be{\begin{equation}}
\def\ee{\end{equation}}

\begin{document}

\title{$K^-$ and $\bar p$ Spectra  for Au+Au Collisions at $\sqrt{s}$ = 200 GeV from \\
        STAR, PHENIX and BRAHMS in Comparison to Core-Corona Model Predictions}

\author{C. Schreiber, K. Werner and J. Aichelin\footnote{invited speaker}}

\address{SUBATECH, Universit\'e de Nantes, EMN, IN2P3/CNRS
\\ 4 rue Alfred Kastler, 44307 Nantes cedex 3, France}
\begin{abstract}
Based on results obtained with event generators we have launched the core-corona model. It describes in a simplified
way but quite successfully the centrality dependence of multiplicity and $<p_t>$ of identified particles observed in heavy-ion reaction at beam energies between $\sqrt{s}$ = 17 GeV and 200 GeV. Also the centrality dependence of the elliptic flow, $v_2$, for all charged and identified particles could be explained in this model. Here we extend this analysis  and study the centrality dependence of single particle spectra of $K^-$ and ${\bar p}$ measured by the PHENIX, STAR and BRAHMS collaborations. We find that also for these particles the  analysis of the spectra in the core-corona model suffers from differences in the data published by the different experimental groups, notably for the pp collisions. As for protons and $K^+$ for each experience the data agree well with the prediction of the core-corona model but the value of the two necessary parameters depends on the experiments.  We show as well that the average momentum as a function of the centrality depends in a very sensitive way on the particle species and may be quite different for particles which have about the same mass. Therefore the idea to interpret this centrality dependence as a consequence of a collective expansion of the system, as done in blast way fits may be premature.
\end{abstract}
\maketitle

\section{Motivations}
There is ample evidence by now that  in heavy-ions collisions at beam energies which can be reached at the colliders at CERN and in Brookhaven a plasma of quarks and gluons is created. Such a state is predicted by 
lattice gauge calculations at high density and/or  temperature.  This plasma is a very short-living state - it lasts less than $10^{-23}$ seconds -  but it is assumed that this time is sufficiently long for reaching equilibrium. Assuming such an early equilibrium, whose origin is still debated, hydrodynamical calculations describe many details of the observables.  It was 
surprising that the multiplicity of identified stable particles in the most central collisions agrees almost perfectly with that expected for a statistical distribution at a freeze-out temperature of around 170 MeV and a small baryon chemical potential. 

For symmetric systems the number of projectile participants equals that of target participants, independent of the centrality. If each particpant contributes the same energy in the center-of-mass system and if the system comes to equilibrium, one does not expect that the multiplicity per participating nucleon varies with centrality.  In the experiments such a variation has been observed, however.  In addition the centrality dependence depends strongly on the particle species. Whereas for $\pi$ this ratio is almost constant, for multi strange baryons this ratio varies by a large factor, a phenomenon which has been dubbed strangeness enhancement.

The basic aussumption in statistical model calculations is that geometry does not play a role and {\it all} nucleons
come to statistical equilibrium, means that all phase space configurations compatibel with the overall quantum numbers become equally probable.  In simulations of the heavy-ion reactions on an event-by-event basis
\cite{Werner:2007bf}, however, it has been observed that this not the case. Nucleons closee to the surface of the interaction region suffer from less collisions
than thos in the center of the reaction and there is a nonnegligibla fraction of nucleons which scatter only once and
therefere they will not come to an equilibrium with their environment. The relative fraction of these surface nucleons
decreases with centrality. 

This observation has motivated the core-corona model in which it is assumed that nucleons which scatter initially only once (corona particles)  are not part of the equilibrated source but produce particles as in pp collisions, whereas all the other come to statistical equilibrium (core particles). Of course this fast transition between core and corona particles is a crude approximation but it allows to define from experimental pp and central AA data the centrality dependence of the different observables. Studies have shown that the present quality of data does not allow for a more refined definition of the transition between core and corona particles.
It has been further verified that the core-corona model describes quantitatively the results of the much more involved EPOS
simulation program. 

In a series of papers \cite{Aichelin:2008mi,Aichelin:2010ed,Aichelin:2010ns} it has been shown that the core-corona model describes quite nicely the centrality dependence of the multiplicity of identified particles,  $<p_t>$ of identified particles, spectra of protons and $K^+$ \cite{Schreiber:2010kh} and even of $v_2$ observed in AuAu and PbPb collisions.
The latter has been considered as a test ground for the shear viscosity needed to describe heavy-ion data in viscous hydrodynamical calculations. The core-corona model describes this data without any reference to a viscosity.
The prediction for the CuCu data are completely determined by the AuAu data and agree with data as far as data have been published.

In this contribution we study the observed single-particle $p_t$ spectra for antiprotons, and $K^-$ at midrapidity. 
These two particles are interesting because:\\
\noindent
a) Antiprotons are created in the reaction, whereas a part of the protons are just shifted towards
midrapidity. It is therefore interesting to see whether the spectra show differences. \\
b)  To search for domains in $p_t$ where the deviations from the core-corona prediction is different for particles and antiparticles and to try to interpret those deviations, if they exist, in physical terms.

\section{Centrality and core-corona fraction}
In order to determine the centrality dependence of the specta we have first of all to know the relative contribution of core and corona particles as a function of the centrality. The core-corona model relies on a single parameter : $f(N_{core})$, the fraction of core nucleons as a function of the centrality. Along with the number of participans, $N_{part}$, it is calculated by a Monte-Carlo simulation based on a Glauber model for hadrons in the nucleus. The parameters of the Glauber distribution are the only freedom of this model. We apply here the EPOS approach. The results are presented in \cite{Schreiber:2010kh}. The STAR \cite{Abelev:2008zk} and PHENIX  \cite{Adler:2003cb} collaborations found other and mutually different values of $N_{part }$ for the same centrality bins. The values of the two collaborations agree within error bars but since this is a completely theoretical quantity  it is not clear why the difference cannot be avoided. Because different $N_{part}$ yield different $f_{core}$ it becomes difficult to compare the different experiments with the same model parameters.
In the core-corona model the centrality dependence of the multiplicity and the average $p_T$ of a given particle species $i$ in a centrality bin containing $N_{part}$ participants is given by :
\bea
M^i(N_{part}) &=& N_{part}.\left[f_{core}(N_{part}).M^i_{core} + (1 - f_{core}(N_{part})).M^i_{corona}\right]\\ \nonumber
<p_T>(N_{part})&=& f_{core}<p_T>^{core}+(1- f_{core})<p_T>^{corona}.
\label{eq:M}
\eea
$M^i_{core}$ is the multiplicity per core participant and $M^i_{corona}$ the multiplicity per corona participant. There are several ways to determinate these two values: one can either calculate them from integrated fits or one can use directly the  published values, which are the results from a fit of a specific form (blast wave model) to the experimental spectra. We chose the latter in the present study: we use the multiplicity measured in pp and divided by a factor of two for $M^i_{corona}$. Then we extract $M^i_{core}$ from the most central multiplicity using eq. 1. If there was no suitable pp data like for  PHENIX, we use instead the most peripheral AA bin and eq. 1 to determine $M^i_{corona}$.
$<p_T>^{corona}$ is the published value of $<p_T>$ in pp collision. $<p_T>^{core}$ is obtained from the above equation
applied to the most central bin.

\section{Results}
\subsection{Centrality dependence of $<p_t>$ of identified particles}
\begin{figure}[h]
\centering
\subfigure[Core - Corona $N_{part}$]
{
\includegraphics[scale=0.5]{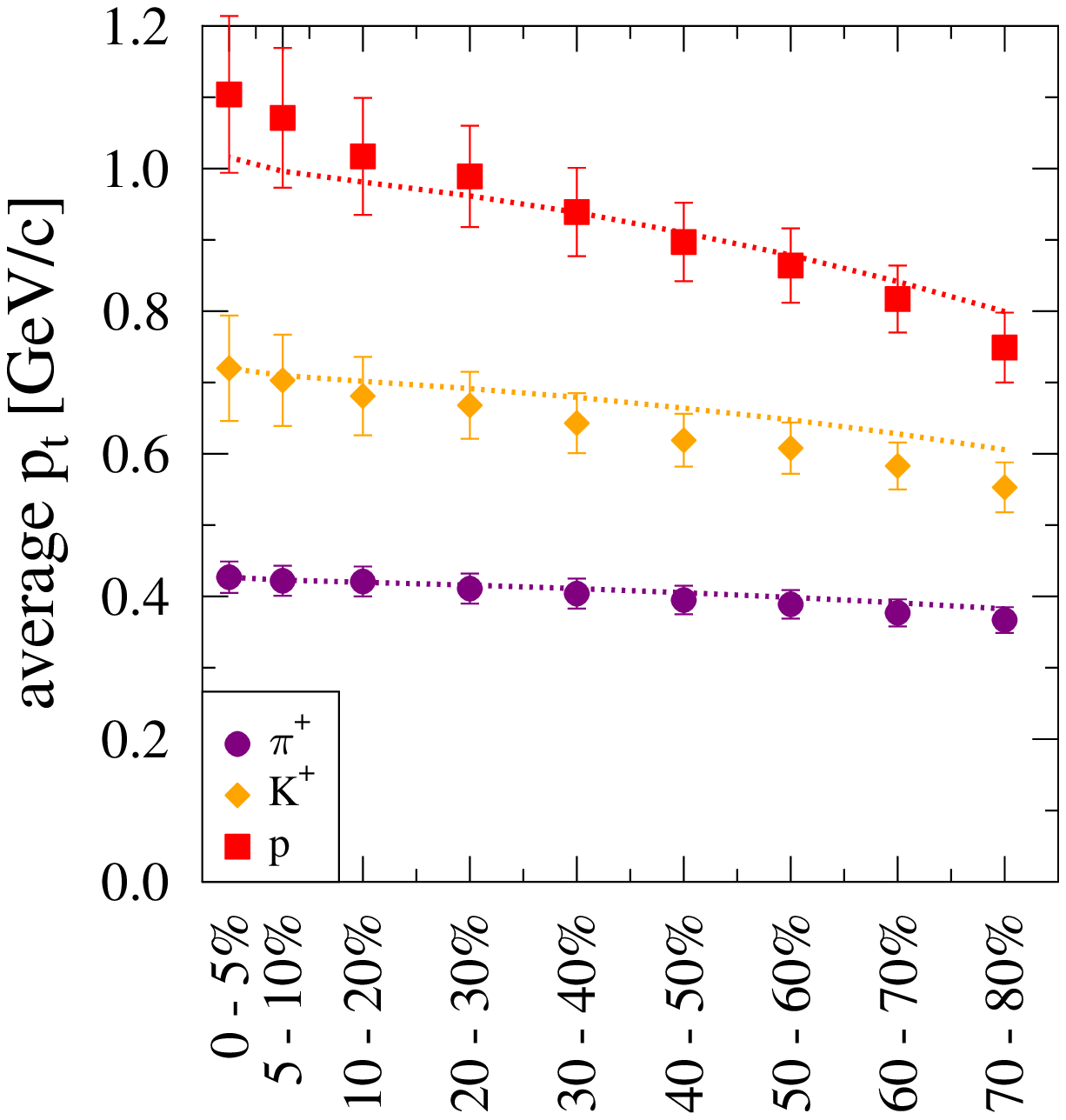}
\label{fig:multCC}
}
{
\includegraphics[scale=0.5]{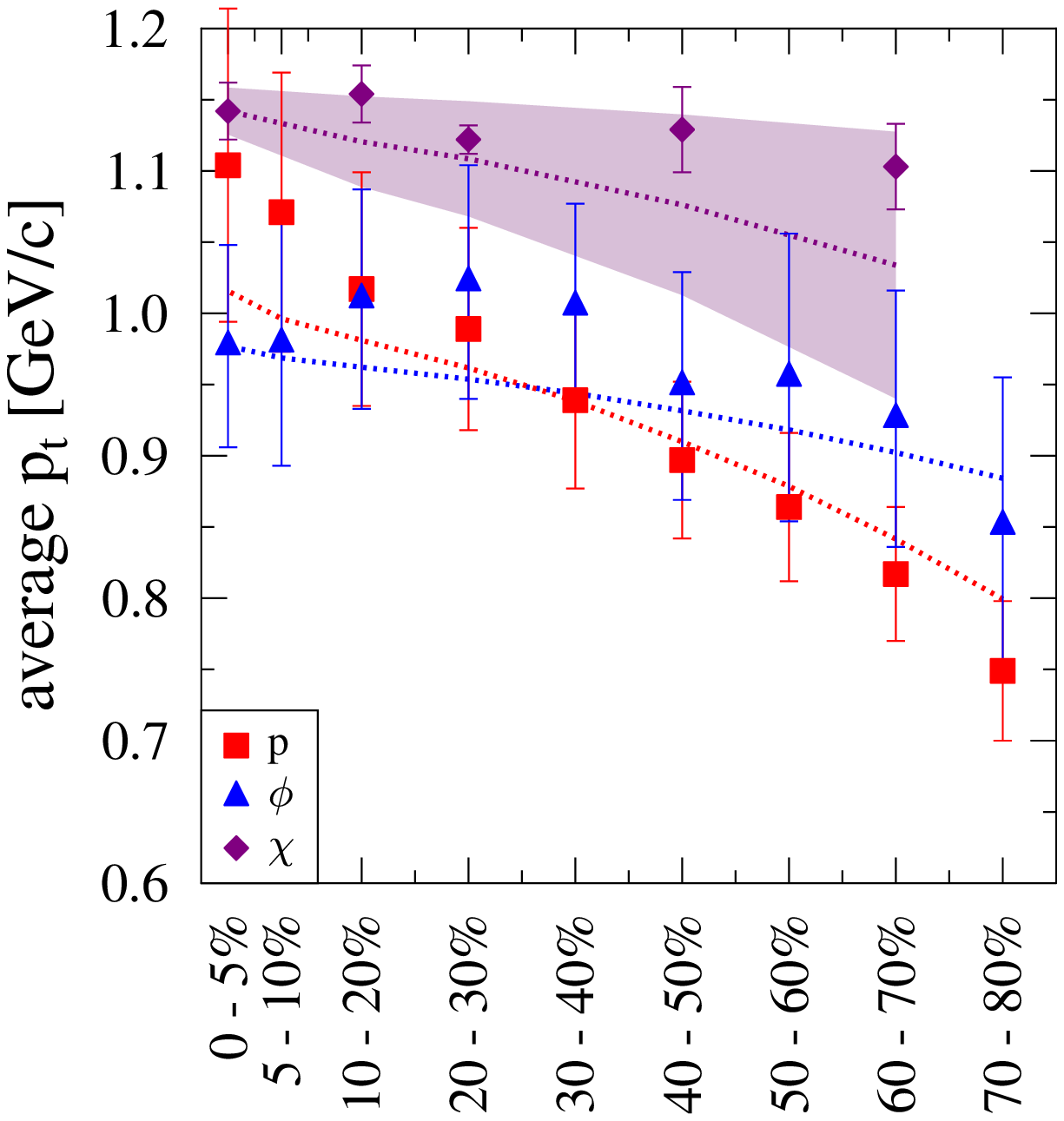}
\label{fig:multResp}
}
\caption{$<p_t>$ as a function of the centrality for different identified particles. On the left hand side we display the
centrality dependence $<p_t>$  for p, $K ^+$, and $\pi^+$ as compared with the results of the core-corona model.
On the right hand side we display the centrality dependence $<p_t>$ for different particles having about the same mass.}
\end{figure}
In this contribution we investigate the centrality dependence of $<p_t>$  for different particles as measured by  the 
STAR \cite{:2008ez}, collaboration. Fig. 1 shows  $<p_t>$ as a function of the centrality for
different identified particles. If we concentrate on  p, $K ^+$, and $\pi^+$,  as shown in fig. 1, left,  we observe
a dependence which can be well described by a blast wave fit in which the increase of  $<p_t>$  with centrality depends on the mass. That the situation is more complicated is displayed in fig. 1, right, which shows that participles with 
about the same mass have a quite different centrality dependence of $<p_t>$. 
The results of the core-corona model are displayed in fig.  1 as lines. The centrality dependence of all particles is well described in this model . For the $\Xi$ we display the theoretical uncertainty of the core corona model by the shaded area.

\subsection{Spectra}
After being formed during the confinement phase transition, the hadrons interact on the way to the detector. 
The results of EPOS  \cite{Werner:2010aa} demonstrate that this rescattering is present but changes the spectra only at low $p_t$, where almost no experimental data are available. EPOS succeeds to reproduce the measured spectra in between a factor of two and, in consequence, to reproduce the change of the spectral form from central to peripheral collisions.  The origin of this success, however, is not very transparent due to the complexity of the approach.  Therefore we decided to study the spectra also in the core-corona model \cite{Aichelin:2008mi,Aichelin:2010ed,Aichelin:2010ns}. In this model we can calculate which spectra would be expected if no final-state interactions among hadrons take place and we can use the difference to the data to learn something about the final-state interaction. Assuming no final-state interactions, in the core-corona model the spectra are superpositions of two contributions: the core contribution and the corona contribution.  The corona distribution  $ \frac{d^2N_i^{corona}}{2\pi . p_t . dp_t . dy}$ is the measured pp spectra divided by two, the core contribution  $ \frac{d^2N_i^{core}}{2\pi . p_t . dp_t . dy}$ is obtained from the experimental spectra for the most central AA collisions corrected for the corona contribution and divided by the number of core participants. Then in the core-corona model the spectra for a given centrality are given by:
\be
\frac{d^2M_i}{2\pi . p_t . dp_t . dy}=
N_{part}[ (1 - f_{core}(N_{part}))\frac{d^2N_i^{corona}}{2\pi . p_t . dp_t . dy}+f_{core}(N_{part})\frac{d^2N_i^{core}}{2\pi . p_t . dp_t . dy}].
\ee
\begin{figure}
\includegraphics[width=0.3\textwidth]{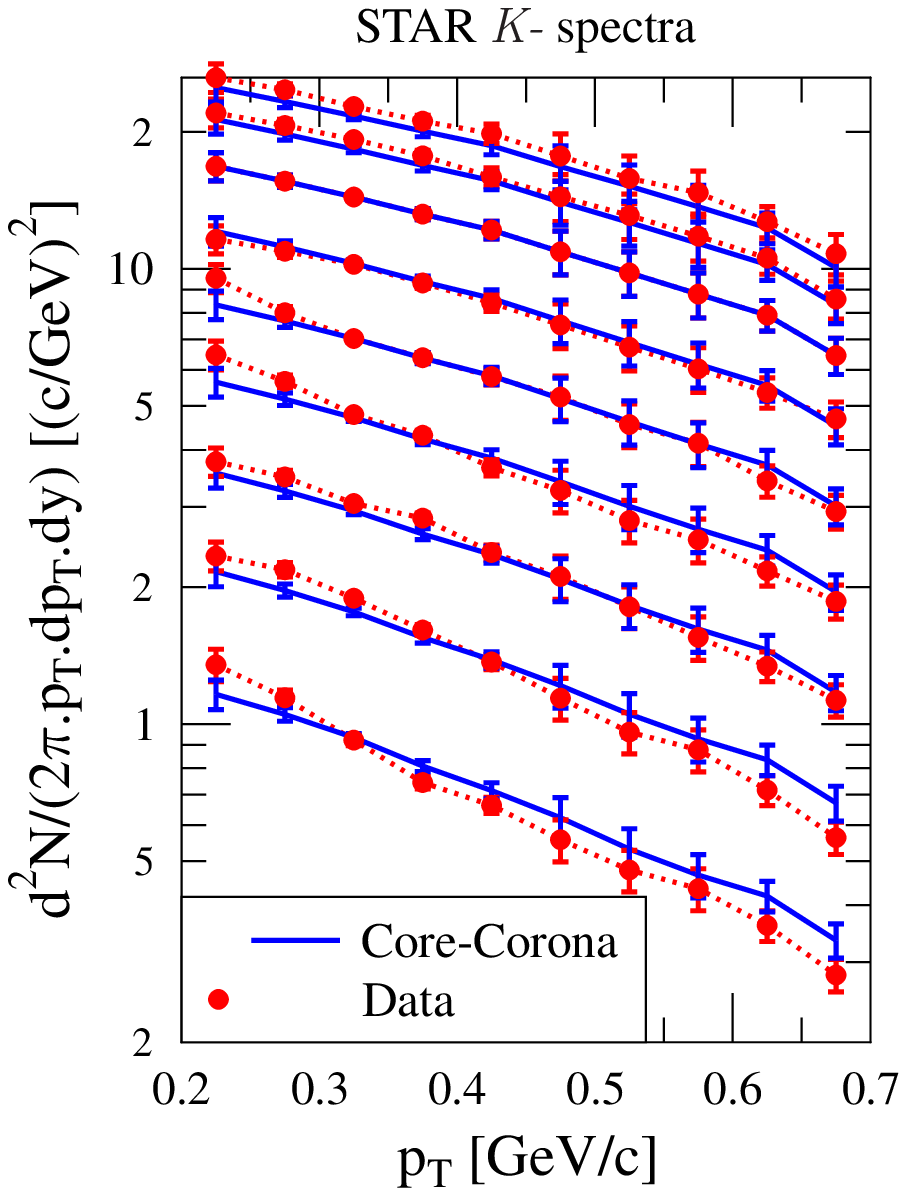}
\includegraphics[width=0.3\textwidth]{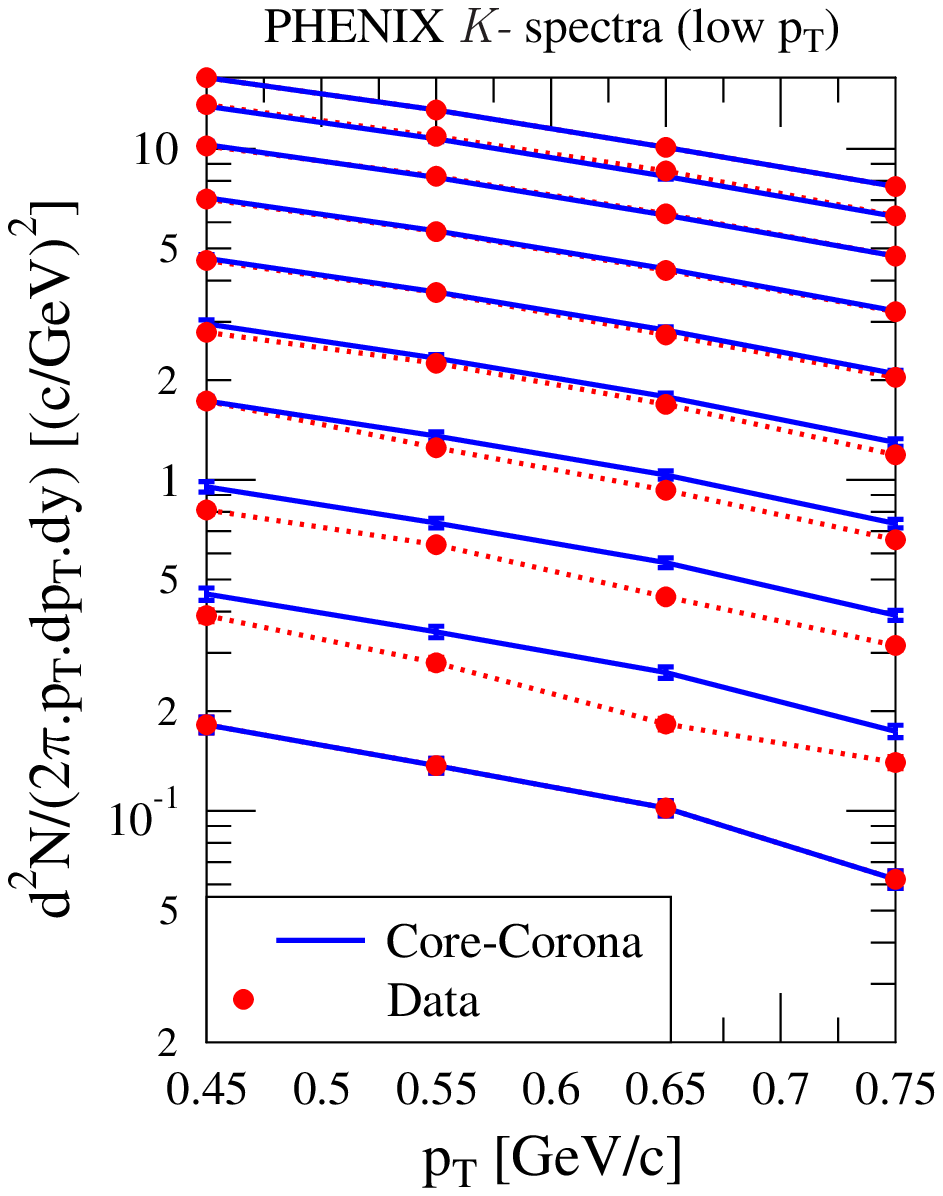}
\includegraphics[width=0.3\textwidth]{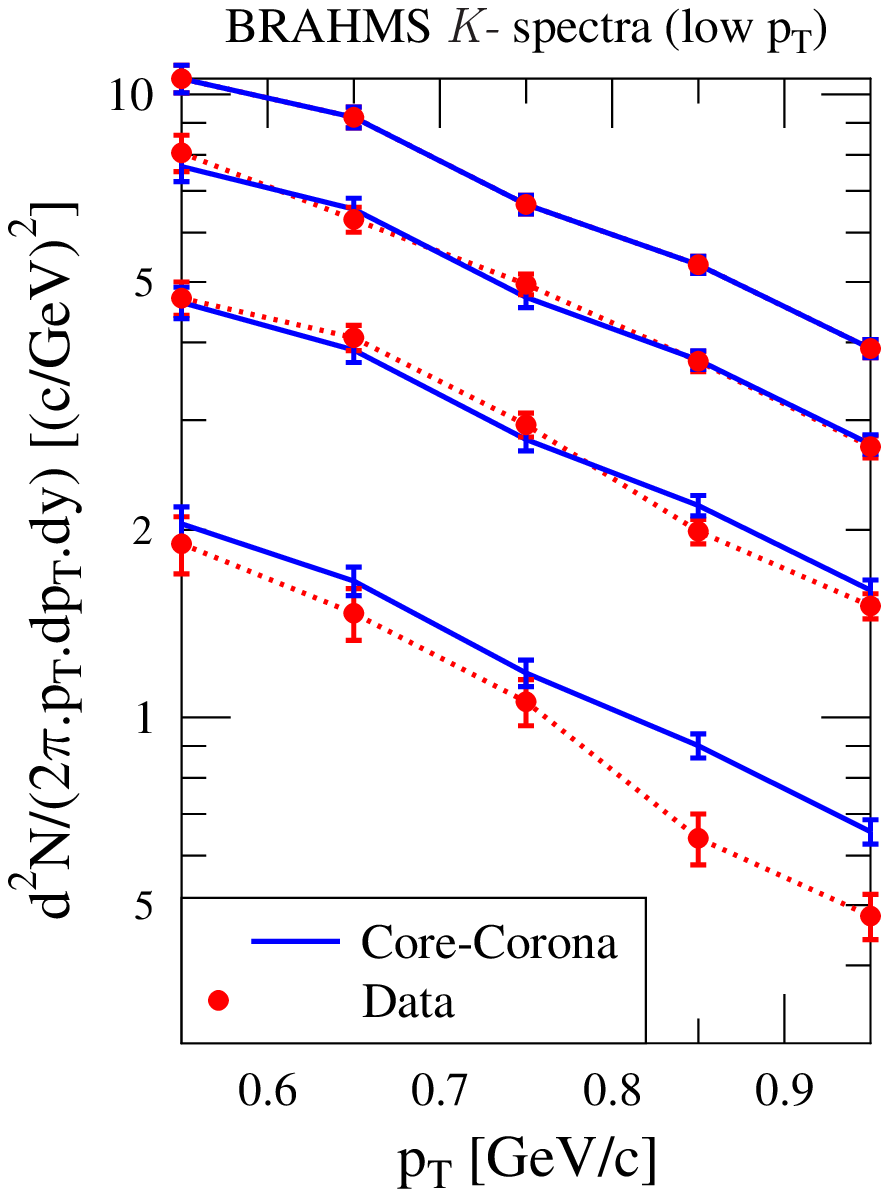}

\includegraphics[width=0.3\textwidth]{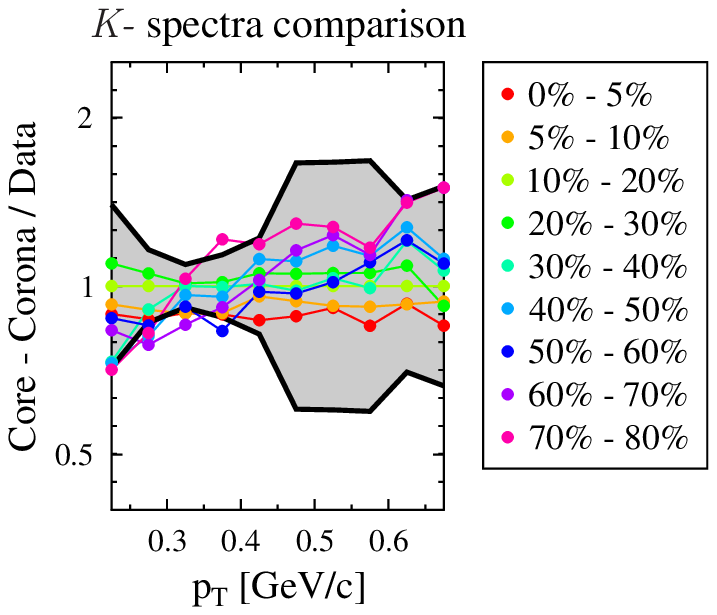}
\includegraphics[width=0.3\textwidth]{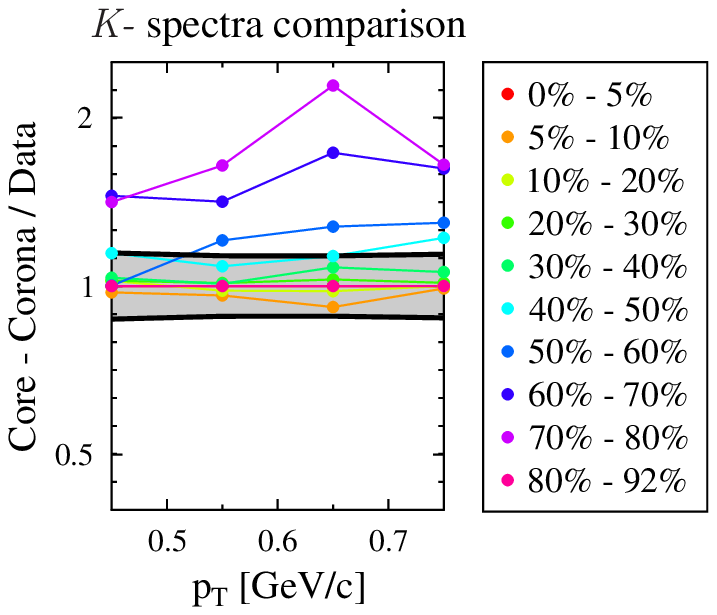}
\includegraphics[width=0.3\textwidth]{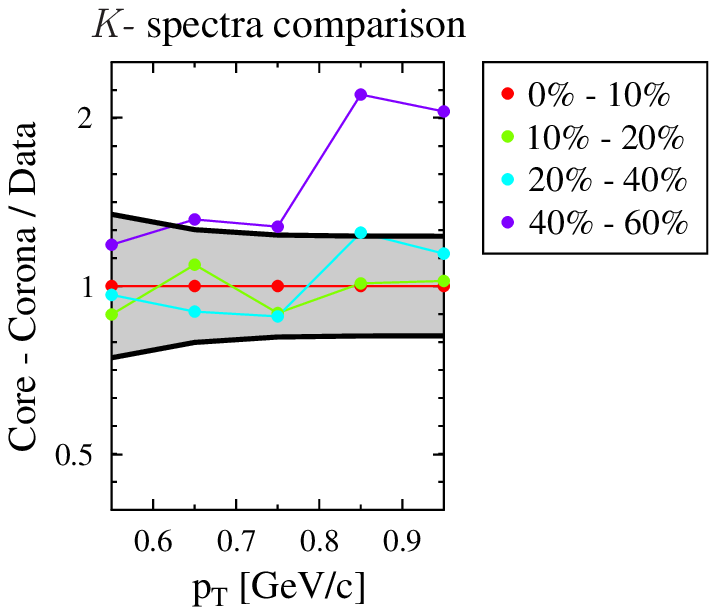}

\includegraphics[width=0.3\textwidth]{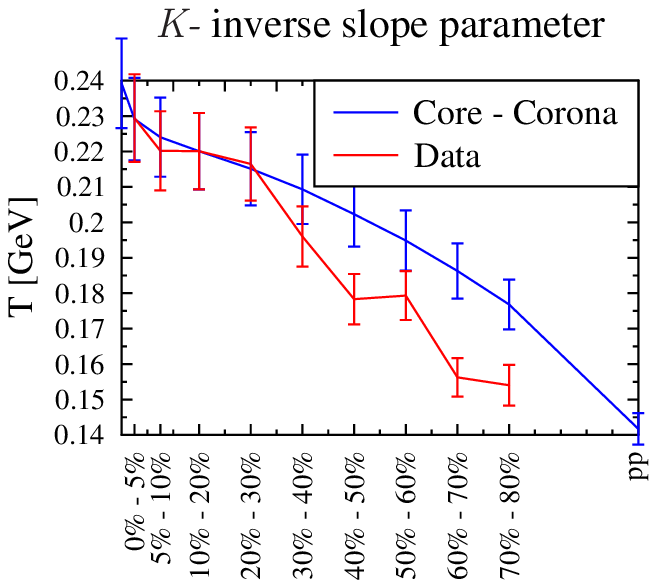}
\includegraphics[width=0.3\textwidth]{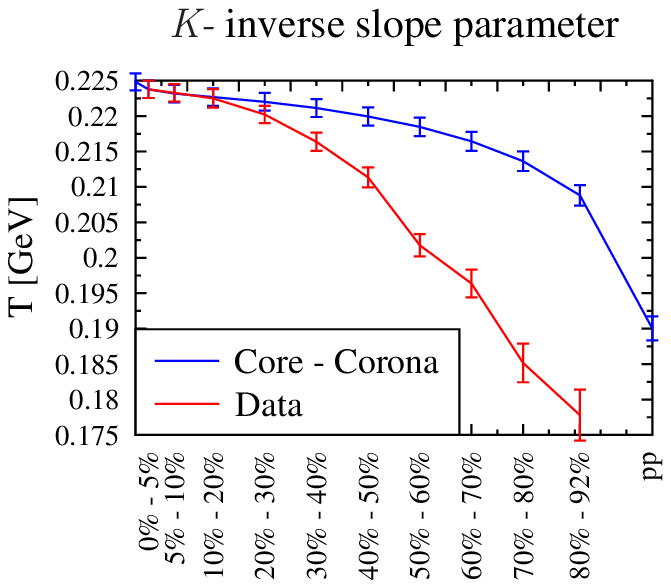}
\includegraphics[width=0.3\textwidth]{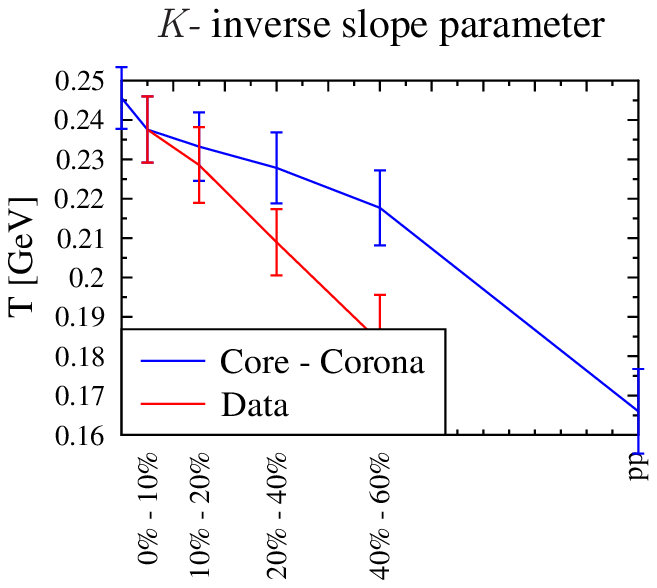}
\caption{Centrality dependence of the $K^-$ spectra measured by the STAR (left), PHENIX (middle) and BRAHMS (right) collaborations in comparison with the predictions of the core-corona model. Top:
The experimental spectra in comparison with the model prediciton. Middle: Ratio of the prediction of the
core-corona model and the experimental data. The shaded area mark the experimental errors. Bottom: Inverse slope parameter T obtained by fitting data and theory by eq. 3. }  
\label{kp}
\end{figure} 
Fig. \ref{kp} presents the results for  $K^-$ mesons. We see in the top row the
spectra measured by the different experimental groups in comparison with the predictions of the core-corona model.
The left hand side presents the STAR data, the middle panel the PHENIX data and  the right hand side the BRAHMS data. 
The middle row displays the difference between theoretical predictions and the data. The shaded region marks the error bars
(which are taken as the averaged error bar over the centrality bins).
We observe that for almost all STAR data points the theoretical predictions are in the experimental error bars. In peripheral reactions there is a tendency that at large $p_t$  the core-corona model is above the data. For semi central reactions model and data are in agreement for all $p_t$ values. This is not trivial at all as the bottom row shows. There we display the inverse slope parameter obtained by fitting the experimental and theoretical spectra by a thermal spectra 
\bea\frac{d^2N}{2\pi.m_t.dm_t.dy} = C.m_t.e^{\frac{-m_t}{T}}
\label{thermo}\eea
with $m_t$ being the transverse mass of the considered particle:
\bea m_t = \sqrt{{p_t}^2 + m^2}\eea
where  $p_t$ is the transverse momentum of the particle (with respect to the beam axis) and $m$ its free mass.
Even if the curves are not exactly exponential and therefore the value of the inverse slope parameters depends on the fit range, the bottom panel shows clearly that the slope varies considerably from central to peripheral reactions (which is in the core corona approach a consequence of the different invariant slope parameters in pp and central AA collisions). Also the PHENIX data are compatible with the core-corona model besides  the second last and third last centrality bin where the deviations, expected from  fig. 1, show up. We note in passing that for those centrality bins the predictions of the core-corona model agree well with the STAR data. The central BRAHMS data are also compatible with the model but we see deviations for the most peripheral bin. It is remarkable that the peripheral STAR data are almost exponential whereas those of PHENIX are not and even less those of the BRAHMS collaboration. Comparing the three experimental spectra with the model we can conclude that for each experiment the majority of data is well described by the model. Deviations are specific for the experiment. There are no systematic deviations.

\begin{figure}
\includegraphics[width=0.3\textwidth]{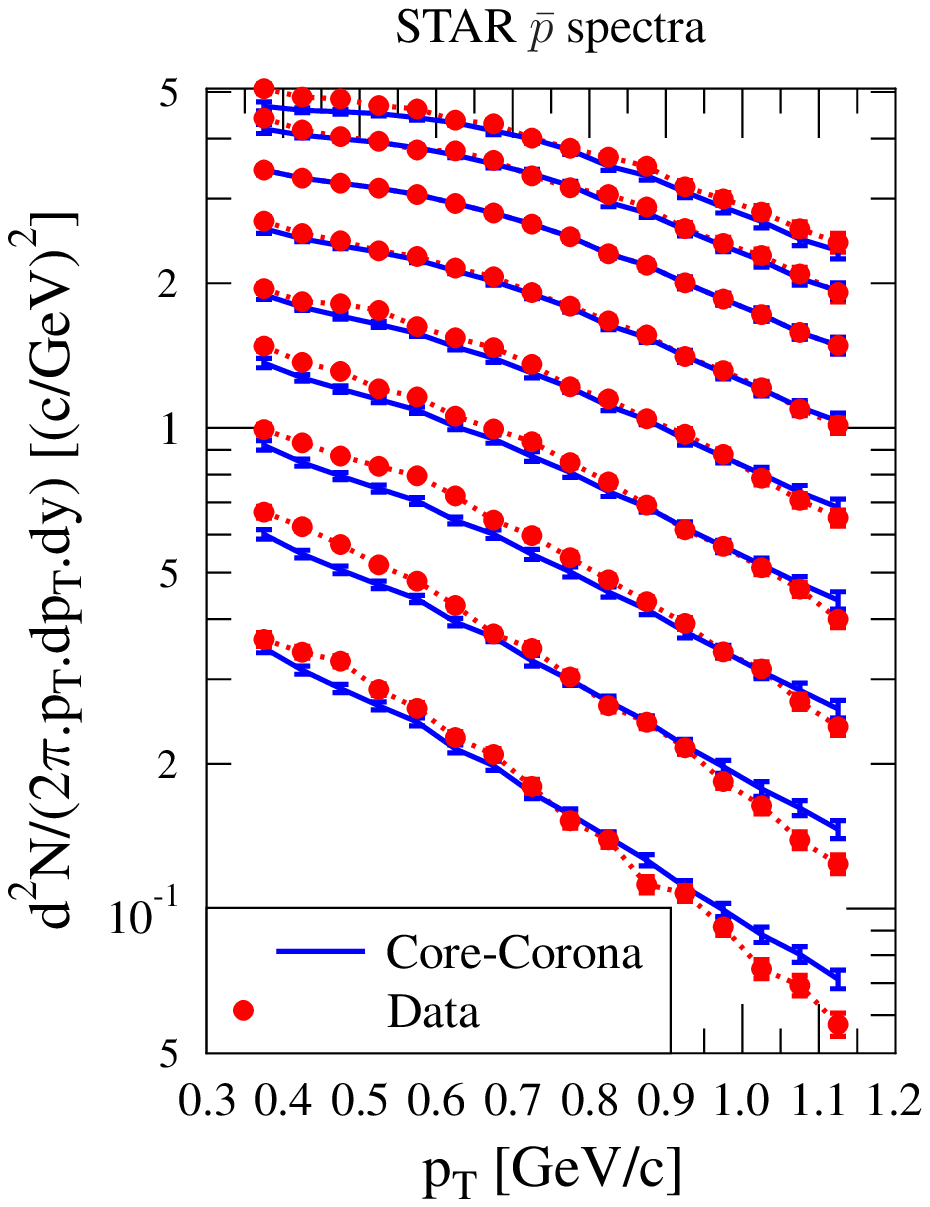}
\includegraphics[width=0.3\textwidth]{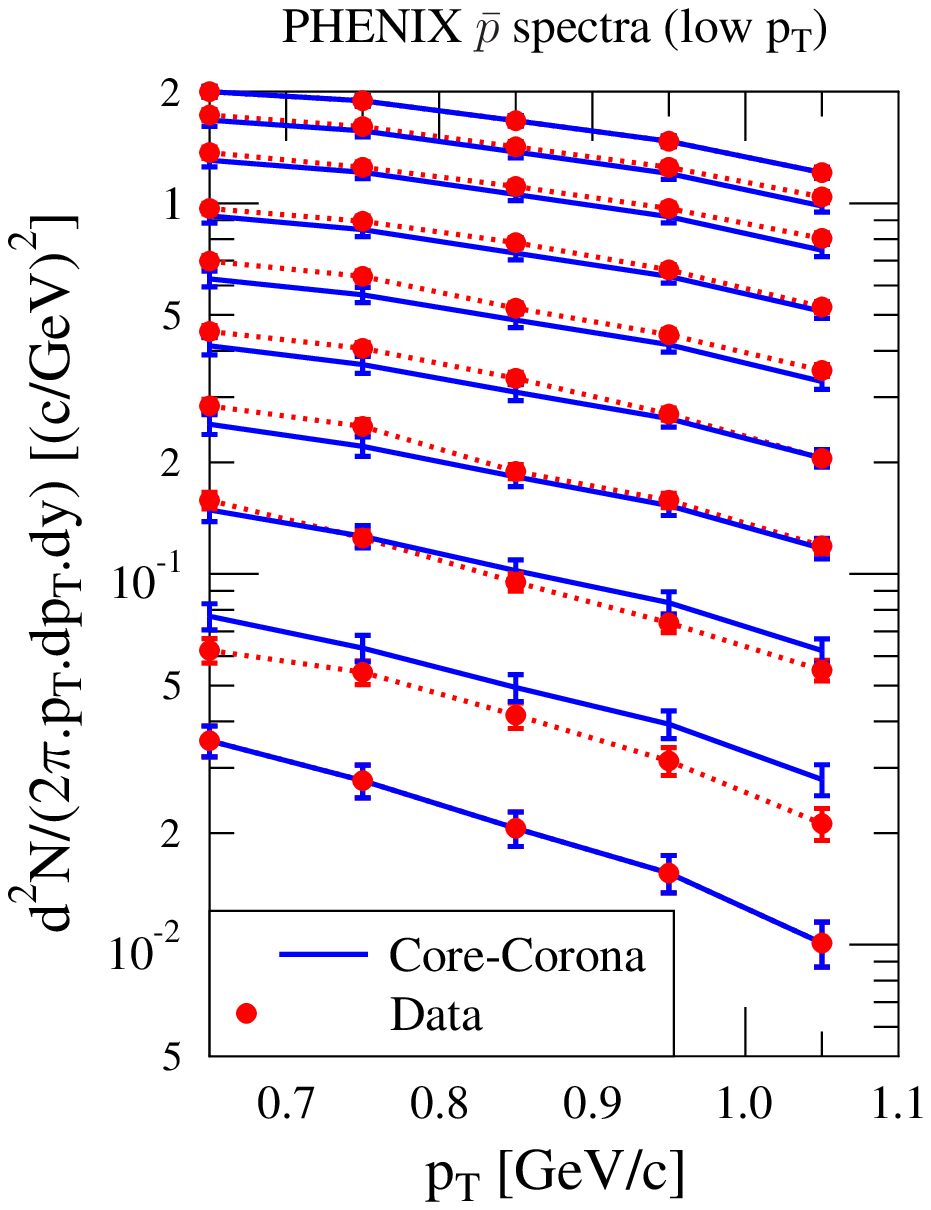}
\includegraphics[width=0.3\textwidth]{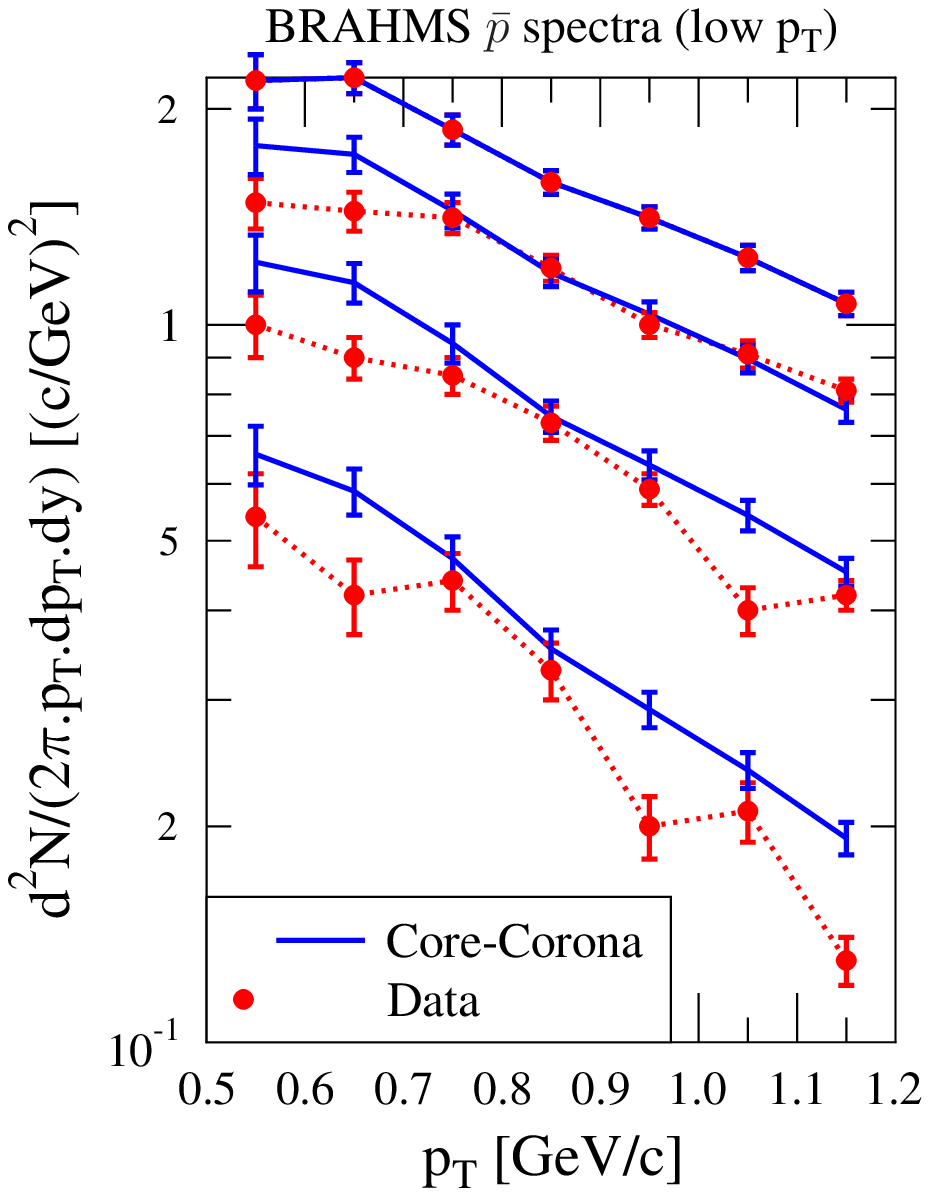}

\includegraphics[width=0.3\textwidth]{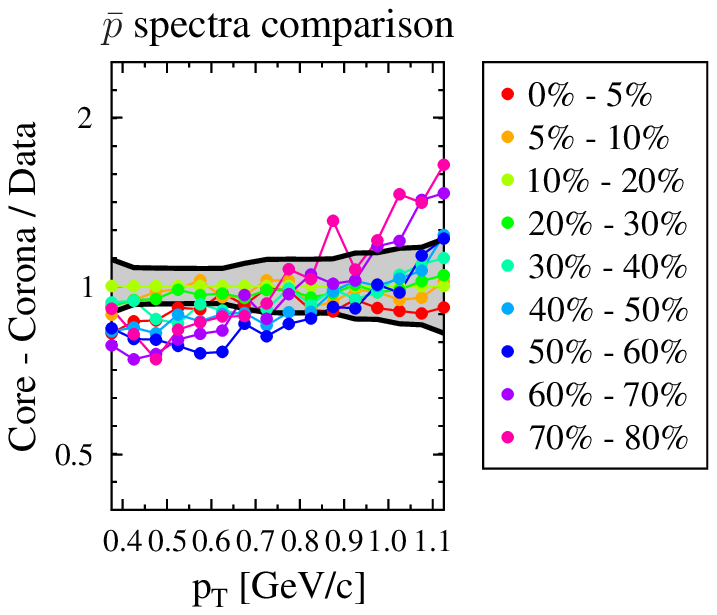}
\includegraphics[width=0.3\textwidth]{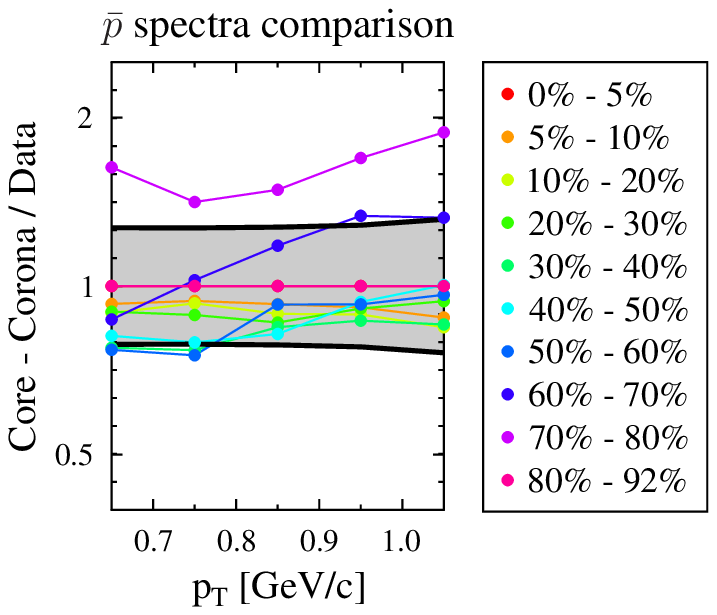}
\includegraphics[width=0.3\textwidth]{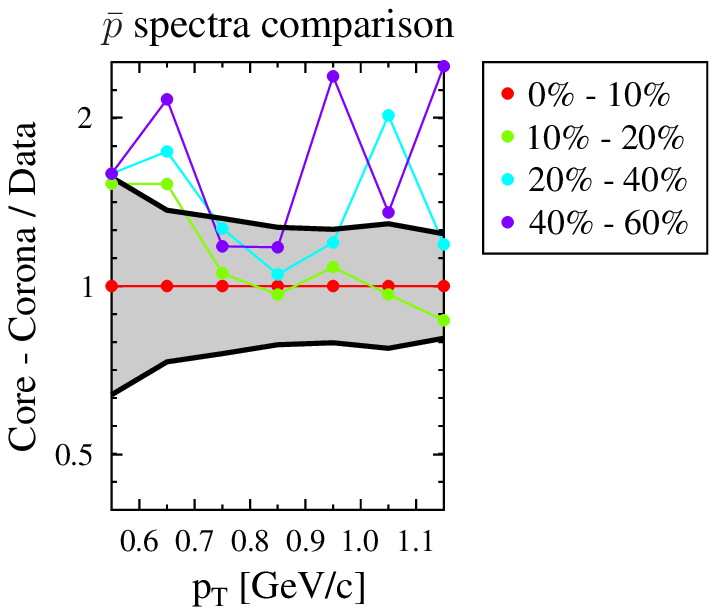}

\includegraphics[width=0.3\textwidth]{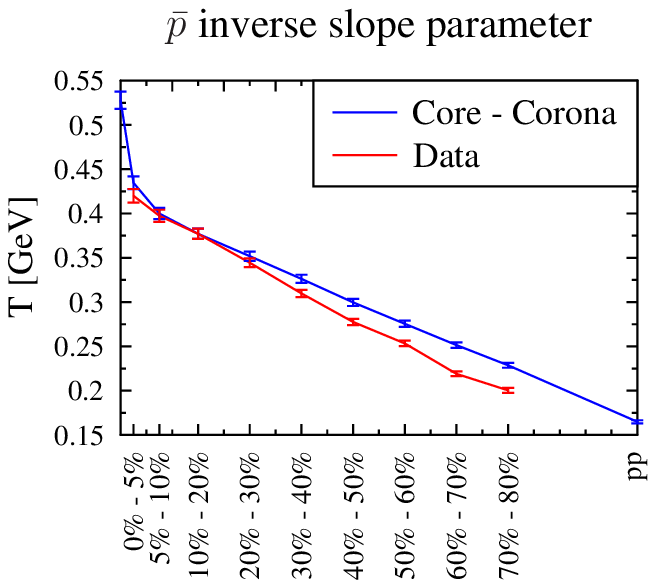}
\includegraphics[width=0.3\textwidth]{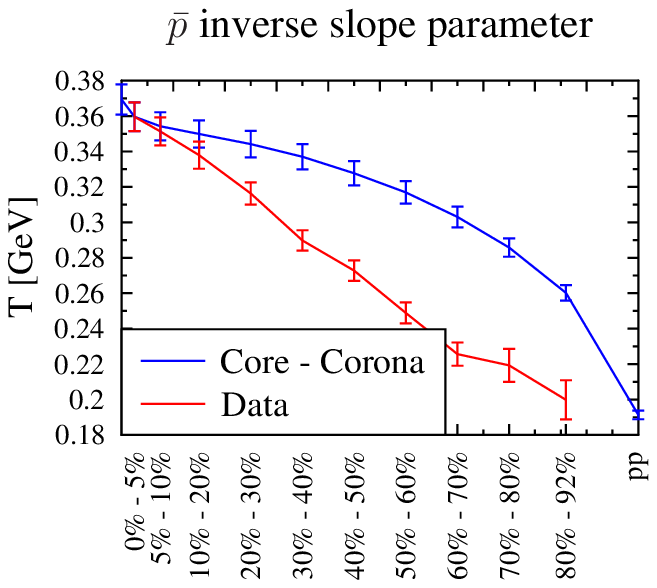}
\includegraphics[width=0.3\textwidth]{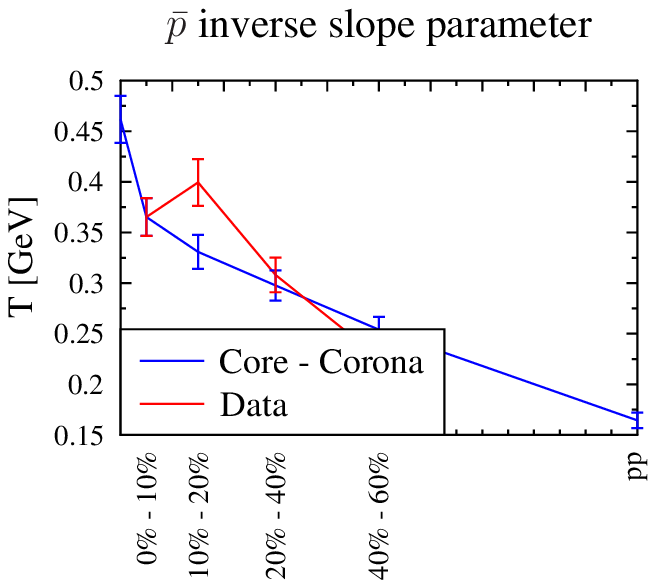}
\caption{Centrality dependence of the antiproton spectra measured by the STAR (left), PHENIX (middle) and BRAHMS (right) collaborations in comparison with the predictions of the core-corona model. Top:
The experimental spectra in comparison with the model prediciton. Middle: Ratio of the prediction of the
core-corona model and the experimental data. The shaded area mark the experimental errors. Bottom: Inverse slope parameter T obtained by fitting data and theory by eq. 3.}  
\label{pp}
\end{figure} 
Fig. \ref{pp} show the same quantities for the antiprotons. For the STAR data see an almost perfect agreement between data and model predictions.
The only exceptions are, as for the $K^-$,  peripheral data at large $p_t$ and the data at $p_t$ around .4-.5 GeV. The middle left figure demonstrates this in detail. We see that  besides the high $p_t$ points in peripheral collisions the predictions of the core-corona model are in between the error bars of the experimental points. This is far from being trivial. The inverse slope parameter of the spectrum varies by a factor of 2 between central
and peripheral reactions and so does the inverse slope parameter of the model due to the large difference between in the inverse slope parameters in pp and central AA. Almost the same is true for the PHENIX data. Here the model overpredicts the data of the second to last centrality bin by an almost constant factor.  In the BRAHMS proton data the form varies from central to peripheral reaction, what is not seen in the STAR and PHENIX data. This can also not be reproduced in the core-corona model

Thus the $K^-$ and $\bar p$ spectra can be described in the core-corona model as well as the $K^+$ and $p$ 
spectra \cite{Schreiber:2010kh}. The spectra of particles and antiparticles are rather similar and the systematic deviations between core-corona prediction and data almost identical. It is, however, impossible to describe the spectra of the different
collaborations by a common model. This is a consequence of the fact that the data of the different collaborations for peripheral reactions are not compatible.

In conclusion, we have shown that all three available data sets for $p_t$ spectra of $\bar p$ and $K^-$ can be well described in the core-corona model. There are deviations but where they occur varies from experiment to experiment. Unfortunately no common parameter set can be found which describes the three experiments simultaneously. This is due to the strong differences between the experimental results for more peripheral events and due to the differences of the spectra measured in pp where the multiplicities differ up to  a factor of two between the three experiments.

Acknowledgement: We would like to thanks Dr. Roehrich and Dr. Yang for making the Brahms pp data available and for an interesting discussion.


\begin{thebibliography}{11}
\bibitem{Werner:2007bf}
  K.~Werner,
  Phys.\ Rev.\ Lett.\  {\bf 98}, 152301 (2007)
  [arXiv:0704.1270 [nucl-th]].
\bibitem{Aichelin:2008mi}
  J.~Aichelin and K.~Werner,
  Phys.\ Rev.\  C {\bf 79} (2009) 064907
  [Erratum-ibid.\  C {\bf 81} (2010) 029902]
  [arXiv:0810.4465 [nucl-th]].
\bibitem{Aichelin:2010ed}
  J.~Aichelin and K.~Werner,
  arXiv:1001.1545 [nucl-th].
\bibitem{Aichelin:2010ns}
  J.~Aichelin and K.~Werner,
J. Phys. G: Nucl. Part. Phys. 37 (2010) 094006.
  arXiv:1008.5351 [nucl-th].
\bibitem{Schreiber:2010kh}
  C.~Schreiber, K.~Werner and J.~Aichelin,
  arXiv:1012.2066 [nucl-th].
\bibitem{Werner:2010aa}
  K.~Werner, I.~Karpenko, T.~Pierog, M.~Bleicher and K.~Mikhailov,
  arXiv:1004.0805 [nucl-th].
\bibitem{Abelev:2008zk}
  B.~I.~Abelev {\it et al.}  [STAR Collaboration],
  Phys.\ Lett.\  B {\bf 673}, 183 (2009)
  [arXiv:0810.4979 [nucl-ex]].
\bibitem{:2008ez}
  B.~I.~Abelev {\it et al.}  [STAR Collaboration],
  Phys.\ Rev.\  C {\bf 79} (2009) 034909
  [arXiv:0808.2041 [nucl-ex]].
\bibitem{Adler:2003cb}
  S.~S.~Adler {\it et al.}  [PHENIX Collaboration],
  Phys.\ Rev.\  C {\bf 69}, 034909 (2004)
  [arXiv:nucl-ex/0307022].
\bibitem{Arsene:2005mr}
  I.~Arsene {\it et al.}  [BRAHMS Collaboration],
  Phys.\ Rev.\  C {\bf 72}, 014908 (2005)
  [arXiv:nucls-ex/0503010].
\bibitem{YangPhD}
  H.~Yang,
  Particle Production in p+p and Au+Au collisions at $\sqrt{s_{NN}}$ = 200 GeV, PhD Thesis, Univeristy of Bergen (Norway) , 2007
\end{thebibliography}
\end{document}